\documentclass[aps,prb,reprint,superscriptaddress]{revtex4-1}

\usepackage{amsmath}
\usepackage{mathtools}
\usepackage{esdiff}
\usepackage{siunitx}
\usepackage{braket}
\usepackage{color}
\usepackage{graphicx}
\usepackage{epstopdf}
\usepackage{soul}
\begin{document}

\title{Cooper pair tunnelling and quasiparticle poisoning in a galvanically isolated superconducting double dot}
\date{\today}

\author{A. A. Esmail}
\author{A. J. Ferguson}
\author{N. J. Lambert}
\email{nl249@cam.ac.uk}
\affiliation{Microelectronics Group, Cavendish Laboratory, University of Cambridge, Cambridge, CB3 0HE, UK}

\begin{abstract}

We increase the isolation of a superconducting double dot from its environment by galvanically isolating it from any electrodes. We probe it using high frequency reflectometry techniques, find $2e$-periodic behaviour, and characterise the energy structure of its charge states. By modelling the response of the device, we determine the quasiparticle poisoning rate and conclude that quasiparticle exchange between the dot and the leads is an important relaxation mechanism.

\end{abstract}

\maketitle

The presence of excess quasiparticle excitations has a deleterous effect on many superconducting technologies, including resonators\cite{Day2003,DeVisser2012}, single photon detectors\cite{Hadfield2009}, on-chip electronic refrigerators\cite{Nguyen2013}, and superconducting qubits\cite{Lutchyn2005,Martinis2009,Catelani2011}. In qubits this is known as quasiparticle poisoning and is often the limiting factor for coherence times. Typically, there is a significant quasiparticle population even at dilution fridge temperature due to incomplete shielding of the qubit from non-equilibrium radiation\cite{Barends2011,Corcoles2011}. Understanding and supression of this is therefore important for successful superconducting technologies.

We have recently found the superconducting double dot (SDD) a useful platform to understand and exploit quasiparticles\cite{Lambert2014c,Lambert2016a,Lambert2017}. The superconducting double dot comprises two superconducting islands, with a charging energy comparable to the superconducting gap, coupled to each other by a Josephson junction. In our previous work, the islands have been connected via tunnel junctions to normal-metal leads, but here we study a galvanically isolated double dot (GIDD). This approach removes one of the key relaxation pathways for the SDD of quasiparticle exchange with the leads\cite{Lambert2014c}, but isolated semiconductor qubits have been shown to have reduced electron temperatures\cite{Rossi2012}. It also prevents quasiparticle poisoning via tunnelling from the leads. Previous studies on a similar system measured via a charge sensor\cite{Shaw2007,Schneiderman} found strictly $e$-periodic behaviour, implying complete poisoning of the device. This was ascribed to back action from the charge sensor. Here we revisit the system, using radio-frequency reflectometry and microwave spectroscopy to assess the poisoning rate. 

\begin{figure}
\centering
\includegraphics{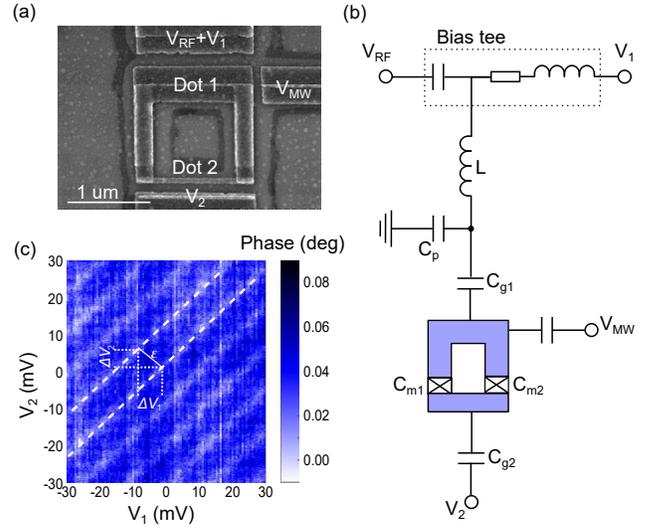}
\caption{(a) An SEM of the galvanically isolated double dot (GIDD). (b) A schematic of the GIDD embedded in the r.f. tank circuit. (c) Phase response of the GIDD as a function of gate voltages $V_1$ and $V_2$ in the normal state ($B=\SI{1}{\tesla}$). The white lines are the gate biases at which electron tunnelling between degenerate charge states occurs.} 
\label{fig1}
\end{figure}

In Fig.~\ref{fig1}(a) we show an SEM of the GIDD, made via double angle shadow mask evaporation\cite{Dolan1977}, with \SI{20}{\nano\metre} thick aluminium forming each island. The SQUID-like geometry allows for the tuning of the Josephson energy of the junction between the two islands with a perpendicular magnetic field. Electrostatic gates allow control of the chemical potentials of the islands via $V_1$ and $V_2$ and the introduction of microwave frequency excitations ($V_{MW}$) and a probe tone for reflectometry ($V_{RF}$). The device is embedded in a tank circuit (Fig.~\ref{fig1}(b)) comprising a lumped element inductor ($L=\SI{560}{\nano\henry}$) and its parasitic capacitance ($C_p=\SI{0.33}{\pico\farad}$), and cooled to a base temperature of \SI{35}{\milli\kelvin}. The inductor has a resonant frequency of $f_{0}=\SI{370.25}{\mega\hertz}$, and, by homodyne detection, we measure the reflected amplitude and phase of a carrier signal at $f_{0}$ and of power -85 dBm at the resonant circuit. This is used to infer changes in the impedance of the device.

In order to suppress quasiparticle poisoning, we take particular care to protect our device from the environment. It is mounted in an enclosure with an infrared absorbent internal coating\cite{Klaassen2002}, the d.c. wiring has microwave (Eccosorb) and low frequency (RC $\pi$ topology) filters, and high frequency cables are attenuated at each temperature stage of the dilution fridge.

We start by applying a large perpendicular magnetic field ($B=\SI{1}{\tesla}$), in order to put the device in the normal state. The number of electrons on each island is quantised, and we label the overall charge state $(m, n)$, where $m$ ($n$) is the charge on dot 1 (2). We measure the strongly averaged phase of the reflected carrier signal as a function of $V_1$ and $V_2$ (Fig.~\ref{fig1}(c)). We observe a series of diagonal lines, corresponding to the gate voltages at which there is a charge degeneracy between the $(m+1,n)$ and $(m,n+1)$ states. At this point, the carrier signal drives incoherent tunnelling between the two islands, resulting in a Sisyphus impedance\cite{Persson2010a,Ciccarelli2011,Lambert2014a} and a corresponding phase shift in the carrier.

Because the overall charge occupancy of the device is fixed, we do not see the honeycomb stability diagram usually observed for double dots\cite{VanDerWiel2002}. It is straightforward to determine gate capacitances for our device ($C_{g1}\approx\SI{27}{\atto\farad}$, $C_{g2}\approx\SI{28}{\atto\farad}$), but to estimate the capacitances for the tunnel junctions we rely on comparison of the area of the junction (as measured from electron micrographs) with previously measured junctions with the same oxide thicknesses\cite{Lambert2014c,Lambert2017}. We determine the capacitance of left junction to be $C_{m1} \approx \SI{310}{\atto\farad}$ and that of the right junction to be $C_{m2} \approx \SI{390}{\atto\farad}$. The total capacitance is therefore $C_{m} = C_{m1} + C_{m2} \approx \SI{700}{\atto\farad}$.

To estimate the resistance of the tunnel junctions, we measure a test junction fabricated at the same time as the device, which has approximately the same junction area and oxide thickness as one of the junctions of the GIDD. We measure a resistance of \SI{66}{\kilo\ohm} at \SI{4}{\kelvin};  the resistance between the two islands of the GIDD, which has two such junctions connected in parallel, is therefore \SI{33}{\kilo\ohm}. From the thickness of the aluminium\cite{Court2007}, we estimate the superconducting gap to be $\Delta = \SI{250}{\micro\electronvolt}$. We then determine a value for the Josephson energy, using the Ambegokar-Baratoff relation, of $E_J = h\Delta/8e^2R = \SI{24}{\micro\electronvolt}$.

\begin{figure}
\centering
\includegraphics{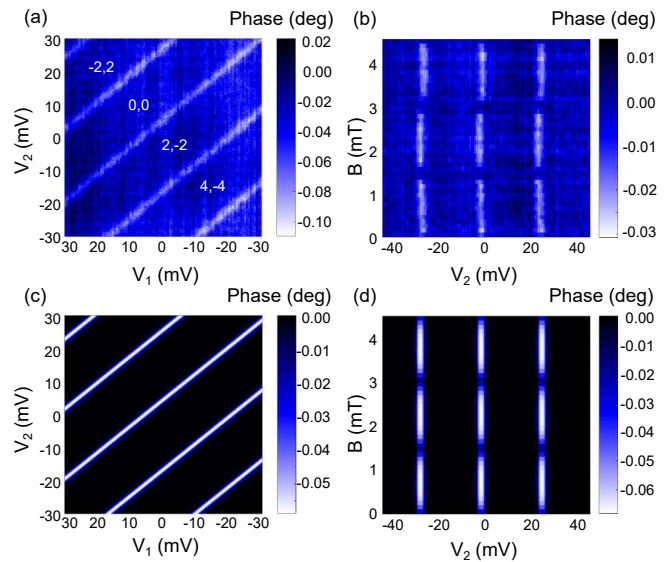}
\caption{(a) Phase shift of carrier signal as function of gate voltages $V_1$ and $V_2$ in the superconducting state. The period of lines indicating the boundaries between charge states halves doubles compared to the normal state (Fig.~\ref{fig1}(c)). (b) Phase shift against gate voltage $V_2$ and out-of-plane magnetic field, $B$ over three gate periods and three flux periods. (c) Simulated phase shift for panel (a), using the estimated device parameters and model in section 3. (d) Simulated phase shift for panel (b).} 
\label{fig3}
\end{figure}

We now reduce the magnetic field to $B=\SI{1}{\milli\tesla}$, and repeat measurements of carrier phase as a function of $V_1$ and $V_2$ (Fig.~\ref{fig3}(a)). We again observe diagonal lines, but with twice the gate period to those seen in the normal state. We therefore conclude that the behaviour of the device is now $2e$ periodic. In this case, the observed change in impedance is the quantum capacitance of the device due to the anticrossing between charge states coupled by the transfer of one Cooper pair between the islands\cite{Petersson2010,Lambert2014c}. 

Next we study the behaviour of the device as a function of perpendicular magnetic field. In Fig.~\ref{fig3}(b) we show carrier phase shift as a function of magnetic field and $V_2$, over three gate periods. The phase shift is also periodic in $B$, with a period of $\Delta B=\SI{1.53}{\milli\tesla}$. This period corresponds to the increase in the flux through the loop by one flux quantum, $\Phi_0$, and for our loop of area $A=\SI{1.1}{\micro\metre}\times\SI{1.2}{\micro\metre}$, $A \Delta B \approx \Phi_0$.

To support our estimates for the device parameters, we model the quantum capacitance of a GIDD as a function of the gate voltages. Quantum capacitance is the polarizability associated with the Cooper pair eigenstates and has been used to readout the quantum state of devices consisting of Josephson junctions\cite{Duty2005,Sillanpaa2005,Lambert2014c}. The quantum capacitance is given by $C_Q = -\partial^2E/\partial V_{g1}^2$, where $E$ correponds to the energy of the occupied eigenstate of the GIDD. The eigenenergies are derived from the Hamiltonian of a superconducting double dot,
\begin{equation}
\begin{aligned}
H &= \sum_{m,n} (U_\textit{E} + \Delta \left (m \bmod{2} + n \bmod{2} \right ) ) \ket{m,n} \bra{m,n}\\
&-\sum_{m,n \text{ even}} \frac{E_J}{2}(\ket{m+2,n}\bra{m,n+2}\\ 
&\qquad\qquad\qquad+ \ket{m,n+2}\bra{m + 2, n} ).
\end{aligned}
\end{equation}
Here, $U_{E}$ is the electrostatic internal energy, which is derived from the capacitance matrix of the device\cite{VanDerWiel2002}. For the galvanically isolated double dot, the source and drain capacitances are set to zero.

We then calculate $C_Q$ and then convert the change in capacitance to a phase shift of the LC circuit. This is done by first calculating the frequency shift due to the change in capacitance, $\delta f = - (f_0/2C_p) \delta C =  (\SI{0.56}{\mega\hertz\per\femto\farad}) \cdot \delta C$, and then the phase shift per unit frequency of the LC circuit near resonance, $\delta \phi / \delta f = \SI{5.6}{\degree\per\mega\hertz}$. 

The calculated phase shift is smoothed using a moving average filter over a 2.5 mV range to simulate the effect of RF averaging (over the voltage window equal to the magnitude of the RF carrier signal) and $1/f$ charge noise\cite{Constantin2009}. We set the temperature $T = 125$ mK, which is the same as the electron temperature determined with a superconducting double dot in a similar environment. In Figs \ref{fig3}(c) and \ref{fig3}(d) we show simulated phase shifts for the experimental conditions in Figs \ref{fig3}(a) and \ref{fig3}(b).

For conventional double dot systems, it is common to define a charging energy, $E_{C1(2)}$, which corresponds to the change in electrochemical potential of dot 1 (2) when an electron is added to dot 1 (2) from an electron reservoir, and an interdot coupling energy, $E_{Cm}$, which is the change in electrochemical potential energy of dot 1 (2) when an electron is added to dot 2 (1) from the reservoir. In the case of the GIDD, these energy scales are not as useful as electrons can only move between the two dots. We find it useful instead to define a transfer energy
\begin{equation}
\begin{aligned}
E_T =& \frac{1}{2} ( \mu_1 (m + n) -  \mu_1 (m, n + 1)\\
&+ \mu_2 (m + 1, n) - \mu_2 (m, n + 1) )\\
=& \frac{E_{C1} + E_{C2}}{2} - E_{Cm}.
\end{aligned}
\end{equation}
This is the change in electrochemical potential (averaged over the two dots) when an electron is transfered from one dot to the other. From our model, we get $E_T=\SI[parse-numbers=false]{112 \pm 20}{\micro\electronvolt}$.

\begin{figure}
\centering
\includegraphics{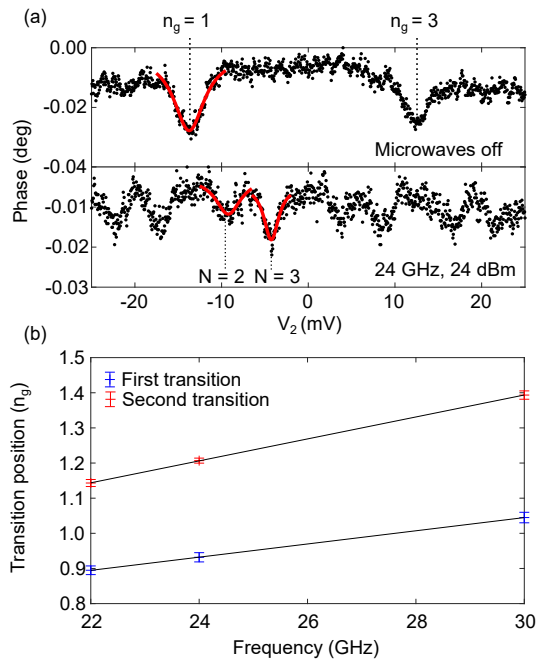}
\caption{(a) Upper panel. Phase at low power, with a fit to one of the peaks due to quantum capacitance. Lower panel. Phase at high power, with fits to peaks corresponding to photon assisted tunnelling processes. (b) Transition distance from quantum capacitance signal at the anticrossing. The ratio of the gradients is $\approx 1.5$, indicating that the visible transitions are the two and three photon transitions.} 
\label{fig5}
\end{figure}

We can also measure this energy using microwave spectroscopy. We vary the incident microwave power at different frequencies and measure the carrier phase as a function of $V_2$, with $V_1$ fixed. At high incident powers (Fig.~\ref{fig5}(a), lower panel), multiple peaks appear offset from the undriven quantum capacitance peak (Fig.~\ref{fig5}(a), upper panel). These correspond to the formation of dressed states - where Cooper pair states hybridise with the photonic states of the microwave drive - and appear as avoided crossings\cite{Wilson2007}. These phase shifts are observed away from the anticrossing for three frequencies, 22, 24 and 30 GHz. Away from the anticrossing (i.e.~at normalised gate charges $n_g \ll n_{odd}$ or $n_g \gg n_{odd}$, where $n_{odd}$ is an odd integer), the energy of the $N$-photon transition is given by\cite{Persson2010a}
\begin{equation}
Nhf \approx 4 E_T \left (n_g - 1 \right ).
\end{equation}
We fit Lorentzians to the peaks to determine their position on the normalised gate charge axis defined along $\varepsilon$ (labelled in Fig.~\ref{fig1}(c)).

In Fig.~\ref{fig5}(b), we plot the transition position (relative to the quantum capacitance signal at $n_g = 1$) against microwave frequency and calculate the gradients $\approx Nh/4E_{T}$. The gradient for the first transition is $18.8 \pm 1.6$ THz$^{-1}$, and for the second transition is $31.2 \pm 3.7$ THz$^{-1}$. The ratio of these gradients is, within experimental error, 2:3, indicating that the peaks correspond to the two and three photon transitions, respectively. We find that $E_T=\SI[parse-numbers=false]{110 \pm 12}{\micro\electronvolt}$, which agrees with our estimate from the device capacitances.

The agreement between microwave spectroscopy measurements and the energy scales deduced from quantum capacitance measurements is excellent up to a difference in the magnitude of the phase shift of a factor of 0.36 (\SI{0.07}{\degree} for the simulation (Fig.~\ref{fig3}(c) and (d)) and \SI{0.025}{\degree} for the measurement (Fig.~\ref{fig5}(a) upper panel)). We ascribe this to a poisoning ratio of the double dot system of $\approx 64 \%$, supressing our time averaged measurements of quantum capacitance by the same amount. Despite improvements in shielding, this compares poorly to our previous experiments\cite{Lambert2014c}, in which the double dot with leads was poisoned $\approx 12 \%$ of the time. Here, we cannot measure poisoning rates directly as previously, as the r.f. carrier couples into the double dot by a much smaller ($\times\sim 10$) capacitance, reducing the signal to noise ratio

In conclusion, we find the poisoning rate in our experiments to be lower than previous studies\cite{Shaw2007,Schneiderman} in which purely $e$-periodic behaviour was observed. This is most likely due to the the improved protection from non-equilibrium radiation in our experiment. However, an entirely isolated superconducting double dot is disadvantageous in comparison to our previous devices. Although removing tunnel junctions to the leads increases the isolation from the environment, quasiparticle poisoning is increased as the key relaxation pathway for quasiparticles is removed. A possible approach to improve poisoning levels in this system is to include small quasiparticle traps for each island\cite{Joyez1994,Riwar2016}. Finally, it would be advantageous to be able to probe the poisoning events in real time\cite{Lambert2014c}; by embedding the GIDD in a superconducting resonator, it should be possible to improve the sensitivity of our measurement in order to access this regime.

\begin{acknowledgments}
We acknowledge the support from Hitachi Cambridge Laboratory and EPSRC Grants No. EP/K027018/1 and EP/K025562. A.J.F. was supported by a Hitachi Research fellowship.
\end{acknowledgments}

\bibliography{ISDD}

\end{document}